\documentclass[12pt]{article}
\usepackage{graphicx}

\def\ignore#1{{}}

\let\oldtheequation=\theequation
\def\doteqs#1{\setcounter{equation}{0}            
\def\theequation{{#1}.\oldtheequation}}
\newcounter{sxn}
\def\sx#1{\addtocounter{sxn}{1} \vskip 1.cm  \goodbreak
\noindent{\large\bf\leftline{\thesxn.~~#1}} \nobreak \vskip -.5cm}
\def\sxn#1{\sx{#1} \doteqs{\thesxn}}

\newcounter{axn}

\date{}

\newdimen\mybaselineskip
\newcommand{\beeq}{\begin{equation}}
\newcommand{\eneq}{\end{equation}}
\newcommand{\beqn}{\begin{eqnarray}}
\newcommand{\eeqn}{\end{eqnarray}}

\def\la{\raise.16ex\hbox{$\langle$}\lower.16ex\hbox{}  }
\def\ra{\, \raise.16ex\hbox{$\rangle$}\lower.16ex\hbox{} }

\def\psibar{ \psi \kern-.65em\raise.6em\hbox{$-$} \lower.6em\hbox{} }
\def\psibarb{ \psi \kern-.65em\raise.6em\hbox{$-$}  }

\begin{document}

\thispagestyle{empty}

\baselineskip=12pt



\vspace*{3.cm}

\begin{center}  
{\LARGE \bf Highly Damped Quasinormal Modes of Generic Single Horizon Black holes}
\end{center}

\baselineskip=14pt

\vspace{3cm}
\begin{center}
{\bf  Ramin G. Daghigh and Gabor Kunstatter}
\end{center}

\centerline{\small \it Physics Department, 
University of Winnipeg, Winnipeg, Manitoba, Canada R3B 2E9}
\centerline{\small \it and}
\centerline{\small \it Winnipeg Institute for Theoretical Physics, Winnipeg, Manitoba}
\vskip 1 cm
\centerline{May 6, 2005}

\vspace{3cm}
\begin{abstract}
\end{abstract}
We calculate analytically the highly damped quasinormal mode spectra of generic single-horizon black holes using the rigorous WKB techniques of Andersson and Howls\cite{Andersson}. We thereby provide a firm foundation for previous analysis, and point out some of their possible limitations. The numerical coefficient in the real part of the highly damped frequency is generically determined by the behavior of coupling of the perturbation to the gravitational field near the origin, as expressed in tortoise coordinates. This fact makes it difficult to understand how the famous $ln(3)$ could be related to the quantum gravitational microstates near the horizon.

\baselineskip=20pt plus 1pt minus 1pt

\newpage

\sxn{Introduction }


The highly damped limit of black hole quasinormal modes (QNMs) has recently attracted a great deal of attention, in large part due to the work of Hod\cite{Hod} and Dreyer\cite{Dreyer}. These authors used different, but closely related, arguments to extract information about the underlying black hole quantum gravitational microstates from the real part of the highly damped QNMs. Their arguments can be summarized loosely as follows: For Schwarzschild black holes the real QNM frequency seems to approach a unique value (independent of the nature and angular momentum of the perturbation) as the imaginary part goes to infinity\cite{Nollert-1}. From this fact, one might conclude, as did Hod\cite{Hod}, that this limiting value constitutes a fundamental oscillation frequency associated with the dynamics of the event horizon.
Dimensional arguments require this frequency to be proportional to the Hawking temperature:
\beeq
\omega = \alpha {k\over \hbar} T_{BH}~,
\label{alpha}
\eneq 
where $\alpha$ is a dimensionless constant\footnote{Note that since the Hawking temperature is proportional to $\hbar$ the right hand side is a purely classical quantity, as required. We will henceforth use units in which $\hbar$ and Boltzman's constant $k$ are unity.}.  

Given the existence of such a frequency, semi-classical arguments require the existence of states in the energy spectrum that are separated by the corresponding energy quantum
\beeq
\Delta E_n = \hbar \omega \Delta n~,
\eneq
where $n$ is the integer labeling the states and $\Delta n=1$. In the large $n$ limit this expression can be integrated to yield
\beeq
n=\int {dE\over \omega} = {1\over \alpha} \int {dE\over T_{BH}} ={1\over \alpha} S_{BH}~.
\eneq
By the first law of black hole thermodynamics, this is proportional to the Bekenstein-Hawking\cite{Bekenstein-1,Hawking-1} entropy, which therefore has states that are equally spaced in the semi-classical limit:
\beeq
S_{BH} = \alpha n  ~.
\label{equally spaced}
\eneq
That the black hole area/entropy is an adiabatic invariant with equally spaced spectrum was first conjectured by Bekenstein\cite{Bekenstein-2} and later by Mukhanov\cite{bm}. The above form of the argument applies to all black holes described by a single dimensionful parameter and assumes only the existence of a fundamental vibration frequency associated with the horizon.

It has often been claimed that an equally spaced spectrum appears to contradict Hawking's prediction that black holes radiate energy with a thermal spectrum. That is, the spacing between the resulting emission frequencies are of the same order as the Hawking temperature, even for macroscopic black holes to which the Hawking's  semi-classical arguments should in principle apply. In this regard it is important to remember that the arguments leading to (\ref{equally spaced}) imply only that the area spectrum of spherically symmetric black holes should contain states that are equally spaced, with separation $\alpha$. This does not necessarily imply that the entire black hole area spectrum is equally spaced. Moreover, this prediction applies only to the case of exact spherical symmetry. It is possible, and in fact very likely, that the addition of further degrees of freedom (e.g. deviations from spherical symmetry), will lead to a sufficient number of new lines in the spectrum so as to make it effectively continuous, at least for macroscopic black holes in which the principle quantum number $n$ is very large, and the Hawking temperature is very small. 

Hod was the first to notice from numerical calculations\cite{Nollert-1} that for real part of the highly damped QNMs of 4-d Schwarzschild black holes, the constant of proportionality $\alpha$ appeared to be $\ln(3)$. This leads to a semi-classical Bekenstein-Hawking entropy spectrum of $\ln(3^n)$ which has a natural statistical mechanical interpretation in terms of a black hole horizon made out of $n$ elements of area, each with three internal degrees of freedom. 

Dreyer\cite{Dreyer} discovered that Hod's arguments appeared to be consistent with the  statistical mechanical black hole entropy that arises in loop quantum gravity\cite{Baez}, and remarkably could be used to independently and consistently fix the elusive Barbero-Immirzi parameter.  Unfortunately a recent recalculation\cite{Lewandowski} of the statistical mechanical entropy of black holes in LQG seems to bring into question the validity of Dreyer's analysis. 

These developments subsequently lead to numerous analytic analysis of the highly damped QNMs of generic black holes. The first was due to Motl\cite{Motl1} who  proved that the coefficient $\alpha$ for 4-D Schwarzschild black holes was indeed precisely $\log(3)$, verifying the conjecture by Hod\cite{Hod}.  Motl and Nietzke\cite{Motl2} were able to prove that the $\ln(3)$ persists for Schwarzschild in all dimensions above three, as argued in \cite{Gabor-prl}. Birmingham and Carlip\cite{carlip1} then showed the existence of an intriguing, but somewhat different, relationship between the highly damped QNMs of the BTZ black hole and its quantum gravitational states. There are by now many analytic calculations of the highly damped QNM spectrum for a large variety of black holes\cite{Natario}. It seems that for spacetime dimensions 4 and up, the intriguing conjectures described above apply at most to single horizon, asymptotically flat black holes.

Most calculations of the highly damped QNMs focus on particular black hole solutions. This includes a recent paper by Tamaki {\it et al}\cite{tamaki}, which start with a general metric,but then specialized to 4-D Schwarzschild and dilaton black holes. More recently, two papers have calculated the highly damped QNMs for  large classes of single horizon black holes within a single formalism. Kettner {\it et al}\cite{Gabor3} analyzed non-minimally coupled scalar fields in the background of black holes in generic 2-d dilaton gravity. In an elegant analysis, Das {\it et al}\cite{Das1} looked at the QNMs of scalar fields in the background of completely general static, spherically symmetric black holes in $d$ dimensions. It turns out that the two classes considered in these papers are closely related. 

The purpose of the present paper is to re-examine the most general class of spherically symmetric, asymptotically flat black holes using the rigorous WKB techniques of Andersson and Howls \cite{Andersson}. We are thereby able to provide a firm foundation for previous analysis, and point out some of their possible limitations. Our general formalism also sheds light on the source of the universality and proposed physical significance of the famous $\ln(3)$ in the real part of the highly damped QNM frequency. We show in fact that the $\ln(3)$ is not generic in principle, despite the fact that in practice it does appear valid for higher dimensional, asymptotically flat black holes even in the presence of other dynamical fields (i.e. not just for higher dimensional Schwarzschild). Moreover, the value of this coefficient is generically determined by the exponent that determines the behavior of the coupling of the perturbation to the gravitational field near the origin, as expressed in tortoise coordinates. Although there may be general arguments that restrict the value of this parameter, it is hard to see how it could be related to the quantum gravitational properties of the horizon.

In Sec. 2, we set up the problem by introducing a general equation for scalar perturbations in a static, spherically symmetric black hole background. In Section 3 we present some general properties of the large damping limit of this equation, and make some general observations about the universality of the resulting frequencies. In Sec. 4, we investigate the global structure of the Stokes and anti-Stokes lines for the highly damped QNMs of a specific, but still very large, class of black holes spacetimes.  In Sec. 5, we calculate the corresponding QNM frequencies using the WKB techniques presented in \cite{Andersson}.  In Sec. 6, we investigate other possibilities for the change of variable, which is introduced in Sec. 4 to obtain suitable behavior of the Stokes and anti-Stokes lines.  A summary and discussion is given in Sec. 7.

\vskip 1cm

\sxn{General Formalism}

The general form of the wave equation that we wish to consider can be suggestively written in 2-dimensional form
\beeq
\partial_\mu\left(\sqrt{-g}h(\phi)g^{\mu\nu}\partial_\nu \psi\right)=\sqrt{-g}V(\phi)\psi~,
\label{matter equation}
\eneq
where $g_{\mu\nu}$ is the two metric on the $(r,t)$ plane, and $\phi$ is a scalar with respect to 2-$d$ coordinate transformations. 
Both the metric and dilaton are assumed static, so that one can find coordinates $(x,t)$ in which $\phi=\phi(x)$ and
 the metric takes the form
\beeq
ds^2=-f(x)dt^2+{1\over g(x)} dx^2~.
\label{metric}
\eneq

Note that we are as yet making no assumptions about the dynamics that give rise to this metric.  The functions $f(x)$, $g(x)$, and $h(x)\equiv h[\phi(x)]$ are completely arbitrary. By further restricting the coordinate system, it is possible to eliminate at most one of these functions, so the system is in fact completely specified by two arbitrary functions. $V(x)$ is determined by the properties of the scalar field. For higher dimensional black holes it contains the centripetal term associated with the angular momentum of the perturbation. 

   In order to restrict to single horizon black hole spacetimes we assume that $h(x)$ is monotonic and vanishes at $x=0$, which is a singular point in the spacetime. Moreover, we assume $f(x)$ and $g(x)$ have simple zeros at the same non-zero $x_h$, the horizon location. Their ratio $H(x)= {f(x)\over g(x)}$ is assumed to be a regular, nowhere vanishing, analytic function of $x$\cite{Das1}.

  At this stage it is useful to make contact with two previous analysis. The first is the calculation of Das {\it et al}\cite{Das1}\footnote{The authors are grateful to S. Das for a very useful discussion explaining the significance of these results.} in which the most general static, spherically symmetric metric in $d$ spacetime dimensions was considered:
\beeq
d\hat{s}^2 = - {f}(r)dt^2+{1\over {g}(r)}dr^2 + r^2 d\Omega^{(n)}~,
\label{higher D}
\eneq 
where $n=d-2$ and $d\Omega^{(n)}$ is the line element on the unit $n$-sphere. By dimensionally reducing the wave equation for a minimally coupled scalar field in this background and making the identifications $x=r$ and $h(x) = r^n $, one obtains precisely (\ref{matter equation}), with
\beeq
V=r^n{l(l+n-1)\over r^2}
\eneq
  Alternatively, one can consider (\ref{matter equation}), as in \cite{Gabor3}, to be the equation describing a scalar field non-minimally coupled to a black hole metric and dilaton in generic vacuum 2-$d$ dilaton gravity. In this case one can choose $\phi=x$ to find that\cite{Gabor1}
\beeq
f(x)=g(x)= J(x)-2GM~,
\eneq
where $M$ is the mass of the black hole and $J(x)$ is determined by the dilaton potential, which is different for different theories.

These two formalisms coincide in the special case that the metric in (\ref{higher D}) describes the higher dimensional Schwarzschild spacetime, for which
\beeq
J(\phi)\propto\phi^{1-1/ n}~.
\eneq

An important observation about (\ref{matter equation}) is that the 2-dimensional D'Alembertion is invariant with respect to conformal transformations of the metric. In particular, if one writes
\beeq
ds^2 = f(x)\left[-dt^2 + \left({dx\over F(x)}\right)^2\right]~,
\eneq
the left hand side of the wave equation depends only on the coupling $h(x)$ and   $F(x)\equiv \sqrt{f(x)g(x)}$:
\beeq
-\frac{h(x)}{F(x)}{\partial^2_t }\psi(x,t)+
{\partial_x} \left[
   h(x)F(x){\partial_x}\psi(x,t)\right]=\sqrt{f\over g} V(x)\psi(x,t)~.
\label{tortoise 1}
\eneq
In terms of the redefined field
\beeq
\psi(x,t) \equiv {\Psi(x)\over \sqrt{F(x)h(x)}}e^{-i\omega t}~,
\eneq
one obtains
\beeq
\frac{d^2\Psi}{dx^2}+R(x)\Psi=0 ~,
\label{schrodinger1}
\eneq 
where
\beeq
R(x)= {\omega^2\over F^2(x)}-U(x)~,
\label{Rx}
\eneq
and the scattering potential between the dilaton and the matter perturbation is given by
\beeq
U(x)={1\over 2}\frac{F''}{F}-\frac{1}{4}\left(\frac{F'}{F}\right)^2 + {1\over 2}\frac{h''}{h}+{1\over 2}\frac{F'}{F}\frac{h'}{h}-{1\over4}\left( \frac{h'}{h}\right)^2 + {\sqrt{f\over g}V(x)\over F h}~,
\label{Ux}
\eneq
with prime denoting differentiation with respect to $x$.   

Equation (\ref{schrodinger1}) has singular points at the horizon $x_h$, where $F(x)$ vanishes but $h(x)$ is assumed regular, and at the origin $x=0$, where $h(x)$ vanishes. Furthermore, the potential $U(x)$ is assumed to vanish as $x\to \infty$  faster than
$1/F^2$. 

For future reference we note that in terms of the tortoise coordinate
\beeq
dz = {dx\over F(x)} ~,
\label{tortoise coord}
\eneq
and a rescaled wave function $\overline{\Psi}=\Psi/\sqrt{F}$, the wave equation (\ref{schrodinger1}) takes the simple form
\beeq
\frac{d^2\overline{\Psi}}{dz^2}+\left( \omega^2-U_h(z) \right)\overline{\Psi}=0 ~,
\label{schrodinger}
\eneq
where 
\beeq
U_h\equiv {1\over2}{h''\over h}-{1\over4}\left({h'\over h}\right)^2+{F\over h}V(x)~,
\label{uh}
\eneq
and the prime here denotes differentiation with respect to $z$.  Under the assumptions we made for $F(x)$, $h(x)$, and $U(x)$, the potential $U_h$ goes to zero at both the horizon ($z\rightarrow -\infty$) and spatial infinity ($z\rightarrow \infty$).  Therefore the asymptotic behavior of the solutions in those regions are
\beeq
\overline{\Psi}(z) \approx \left\{ \begin{array}{ll}
                   e^{-i\omega z}  & \mbox{as $z \rightarrow -\infty$ $(x\rightarrow x_h)$}~,\\
                   e^{+i\omega z}  & \mbox{as $z\rightarrow \infty$ $(x\rightarrow \infty)$~.}
                   \end{array}
           \right.        
\label{asymptotic}
\eneq

\sxn{Large Damping Limit and Universality}

Equation (\ref{schrodinger1}) with boundary conditions (\ref{asymptotic}) defines the general problem of finding QNMs for a generic black hole. 
We now consider the QNMs in the large damping limit
\beeq
|\omega^2|\to [{\rm Im}(\omega)]^2 \to \infty~.
\eneq
Since we will be using complex analytic techniques, the behavior of $U(x)$ on the entire complex plane may be relevant.
However, in the large damping limit case, the $\omega^2/F^2$ term in $R(x)$ will dominate $U(x)$ everywhere, unless one of the terms in $U(x)$ diverges. According to our assumptions, this can only happen at the origin, or at the horizon. However, since the $\omega/F^2$ also diverges at the horizon, it will dominate there as well in the large damping limit. Thus, $U(x)$ is essentially irrelevant in this limit except near the origin. Suppose without loss of generality that $h\to x^{\alpha}$, $F \to x^{\beta}$, and $V \to x^{\gamma}$ in this limit. As long as $\alpha+\beta-\gamma<2$, which is the case for all previously considered calculations, $V(x)$ can be neglected.  Thus, for the highly damped QNMs, $R(x)$ can be approximated on the entire complex plane by
\beeq
R(x)\sim \frac{\omega^2}{F^2}-\frac{(\alpha+\beta)[(\alpha+\beta)-2]}{4x^2}~.
\label{Rx2}
\eneq
Note that in the WKB analysis where the two solutions to Eq. (\ref{schrodinger1}) are
\beeq
\left\{ \begin{array}{ll}
                   \Psi_1^{(t)}(x)=Q^{-1/2}(x)\exp \left[+i\int_{t}^xQ(x')dx'\right]~,\\
                   \\
                   \Psi_2^{(t)}(x)=Q^{-1/2}(x)\exp \left[-i\int_{t}^xQ(x')dx'\right]~,
                   \end{array}
           \right.        
\label{WKB}
\eneq
we need to deal with $Q$ rather than $R$.  Here $Q^2=R+${\it extra term}, where the {\it extra term} is to make sure that the WKB solutions (\ref{WKB}) have the appropriate behavior near the origin of the complex $x$-plane.  From Eq. (\ref{Rx2}) we have $R \sim  -(\alpha+\beta)[(\alpha+\beta)-2]/4x^2$ when $x\rightarrow 0$. Using Eq. (\ref{schrodinger1}), it is easy to show that $\Psi \sim x^{1/2 \pm (\alpha+\beta-1)/2}$ as $x \rightarrow 0$.   Therefore, we need to take $Q^2=R-1/4x^2$, so that the WKB solutions $\Psi_{1,2}$ have the correct behavior of the form $\Psi \sim x^{1/2 \pm (\alpha+\beta-1)/2}$ close to the origin. As a result, the approximate behavior of $Q^2$ on the entire complex plane is  
\beeq
Q^2=R(x)-\frac{1}{4x^2}\sim \frac{\omega^2}{F^2}-\frac{(1-\beta)^2(\frac{\alpha}{1-\beta}-1)^2}{4x^2}~.
\label{Q^2-general}
\eneq

Equation (\ref{Q^2-general}) is the key to understanding the universality of the highly damped QNM spectrum for generic single horizon black holes. The analytic WKB techniques used in the present paper allow the QNMs to be determined purely by the structure of the poles at $x=0$ and $x=x_h$. In fact, as we will show, when the calculation can be done using these techniques then the QNMs are determined by the relationship
\beeq
e^{2\pi\omega\over\kappa}=-(1+2\cos((a-1)\pi))~,
\label{generic answer}
\eneq 
where
\beeq
\kappa = {1\over 2} {F'\over F}
\eneq
 is the surface gravity of the black hole and have defined the parameter
\beeq
a = {\alpha\over (1-\beta)}~.
\eneq 
The parameter $a$ is in fact the precise combination of $\alpha$ and $\beta$ that is invariant under coordinate transformations of the form $x\to \tilde{x}=x^q$. In fact, it is easy to verify that $a$ gives the rate at which the coupling $h(x)$ approaches zero as a function of the tortoise coordinate $z$. That is:
\beeq
h[x(z)]\to z^a
\eneq
as $x\to 0$. 

At a heuristic level, one can see that only $a$ and the surface gravity can affect the values of the highly damped QNMs by  noting that in tortoise coordinates $U_h\to {a(a-2)\over 4 z^2}$. Thus for large $|\omega|^2$, the equation has a universal form with different black hole backgrounds being distinguished only by the residues of the poles at $x=0$ and $x=x_h$, i.e. $a$ and $\kappa$ respectively. Moreover, the parameter $a$ is determined by the coupling of the perturbation to the metric, and does not seem to have any direct relationship to properties of the horizon. 

We close this section by noting that $a=1$ for scalar fields in the background of the Schwarzschild metric in $D=n+2$ spacetime dimensions. Since $a$ is coordinate invariant, we will work with the radial coordinate $r$. In this case:
\beeq
F(r)= (1-2GM/r^{D-3})
\eneq   
where $M$ is the ADM  mass as measured at infinity.  Moreover, $h(r) = r^{D-2}$ so that $\alpha = D-2$ and $\beta = 3-D$ to give
$a = (D-2)/(1-(3-D)) = 1$. This in turn leads to the known result
\beeq
{\omega\over T_{BH}}
 = (2n+1) \pi i + ln(3)~.
\eneq

It is interesting to note that $a=1$ also for stringy black holes in 4 and 5 dimensions 
(see \cite{Das1}, in which $a=qD/2$). It seems important to understand the necessary and sufficient conditions on asymptotically flat single horizon black holes which lead to $a=1$. These conditions, coupled with the analysis given above, would account for the universality of the highly damped corresponding QNMs for single horizon black holes. They would also presumably give further insight as to whether the $\ln(3)$ can be related to the underlying quantum gravity theory.
 
\sxn{2-$d$ dilaton gravity with power-law potentials} 
 
 So far we have kept the arguments as general as possible.  
For concreteness, we now focus on a generic class of black holes in 2-dimensional dilatonic gravity. In the conclusions we explain the circumstances under which the techniques and results in the present section can be applied to a much more general setting. 

With the gauge choice of $\phi=x$, and the ``power-law potentials'' considered in \cite{Gabor3}, we have 
\beeq
J(x) = x^{1-b} ~.
\label{Jlimited}
\eneq
The power-law metric is
\beeq
ds^2=-f(x)dt^2+\frac{dx^2}{f(x)}~,
\label{metriclimited}
\eneq
where
\beeq
f(x)=J(x)-2GM~,
\label{fxlimited}
\eneq
which locates the black hole horizon at
\beeq
x_h=(2GM)^{1/1-b}~.
\label{horizon2}
\eneq
We further need to restrict consideration to the sub-class of ``power-law potentials'' for which
\beeq
0<b<1~.
\label{b-range}
\eneq
The upper bound is necessary for the existence of black hole solutions, while the lower bound guarantees that the physically relevant solutions of the action of the 2-$d$ dilatonic gravity are asymptotically flat.

We also need to know the behavior of the coupling parameter $h(x)$ between the dilaton and matter field in the complex $x$-plane.  In the physically motivated case of spherically reduced Einstein gravity the natural choice for the coupling $h$ is $h=x$\cite{Gabor2}.  Therefore it would be reasonable to use $h=x^a$, where $a$ is an arbitrary parameter.  Substituting $h=x^a$ and $V(x)=0$ in the scattering potential (\ref{Ux}), and keeping only the terms which diverge faster as $x\rightarrow 0$ we get   
\beeq
U(x)=\frac{a(a-2)}{4x^2}~.
\label{Ux-powerlaw}
\eneq
Note that $\beta =0$ in this parameterization.

To calculate the QNM frequencies in the highly damped limit where Im$(\omega) \rightarrow -\infty$, we will follow the analytical method based on WKB approximation suggested by Andersson and Howls in Ref.~\cite{Andersson}, which ultimately provides the motivation for the somewhat simpler, monodromy method of Refs.~\cite{Motl1} and \cite{Motl2} that was applied in Refs.~\cite{Gabor3} and \cite{Das1}.

\begin{figure}[tb]
\begin{center}
\includegraphics[height=7cm]{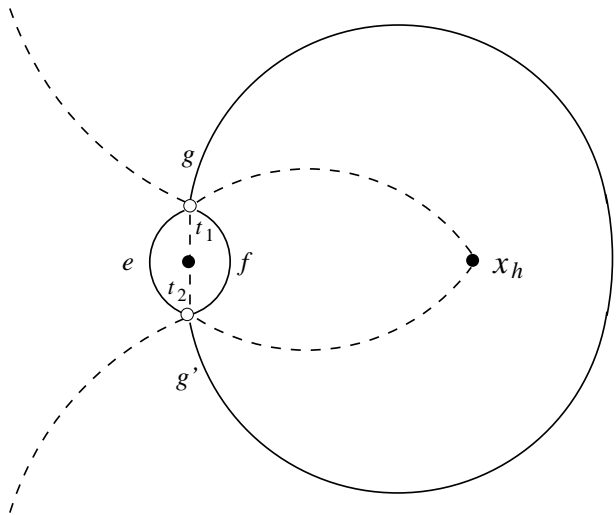}
\end{center}
\caption{A schematic illustration of Stokes (dashed) and anti-Stokes (solid) lines in the complex x-plane.  The hollow circles represent the two zeros of $Q^2$, while the filled circles are the two poles at the origin and the event horizon.}
\label{stokes}
\end{figure}

The zeros and poles of the function $Q$ in the WKB solutions (\ref{WKB}) play a crucial role in determining the behavior of the Stokes lines on which $\int Q dx$ is purely imaginary and anti-Stokes lines on which $\int Q dx$ is purely real.  These lines are vital for the WKB analysis of Eq. (\ref{schrodinger1}).  From each zero three Stokes lines and three anti-Stokes lines emanate.  The angle between Stokes (anti-Stokes) lines is $120^{\circ}$, and the angle between Stokes and anti-Stokes lines is $60^{\circ}$.  The poles of the function $Q^2$ are located at the origin $x=0$, and the event horizon $x=x_h$.   
It is not difficult to show that the zeros of $Q^2$ approach the origin of the complex $x$-plane when Im$(\omega) \rightarrow -\infty$.  In this limit, $Q^2$ may be simplified by expanding Eq. (\ref{Rx}), with $U(x)$ to be the scattering potential (\ref{Ux-powerlaw}), in a power series near $x=0$;
\ignore{\beeq
R \approx \frac{1}{(2GM)^2}\left[\omega^2-\frac{a(a-2)(2GM)^2}{4x^2}\right] \sim -\frac{a(a-2)}{4x^2}~,
\label{asymptotR}
\eneq
when $x \rightarrow 0$.
Using Eq. (\ref{schrodinger1}), it is easy to show that
\beeq
\Psi \sim x^{\frac{1}{2}\pm \frac{\sqrt{a(a-2)+1}}{2}} ~~\mbox{as $x\rightarrow 0$}~.
\label{correctionQ1}
\eneq
On the other hand, if we take $Q^2=R$ we get
\beeq
Q^{-1/2} \sim x^{1/2}~,
\label{correctionQ2}
\eneq
\beeq
\int Q dx \sim \pm i \frac{\sqrt{a(a-2)}}{2}\ln x~,
\label{correctionQ3}
\eneq
which will result in
\beeq
f_{1,2}\sim x^{\frac{1}{2}\pm \frac{\sqrt{a(a-2)}}{2}}~.
\label{correctionQ4}
\eneq
This shows that the approximate solutions do not have the correct behavior in the vicinity of the origin.  This may be corrected by choosing
\beeq
Q^2=R-\frac{1}{4x^2}~.
\label{correctionQ5}
\eneq
The corrected version of Eq. (\ref{asymptotR}) may be written
}
\beeq
Q^2= R-\frac{1}{4x^2} \approx \frac{\omega^2}{f^2}-\frac{(a-1)^2}{4x^2}~.
\label{Q^2}
\eneq
Note that since in this model $h=x^a$, and $f(x)$ approaches to a constant when $x\rightarrow 0$ we can derive this equation by simply substituting $\beta=0$ in Eq. (\ref{Q^2-general}).  It is clear from Eq. (\ref{Q^2}) that the function $Q^2$ has only two simple zeros located on the imaginary axis of the complex $x$-plane.  The schematic behavior of the Stokes and anti-Stokes lines on the entire complex plane is displayed in Fig.~\ref{stokes}.  As one can see, there are no anti-Stokes lines extending to infinity.  The unbounded anti-Stokes lines are crucial ingredients in solving for QNM frequencies in the analytical methods suggested in Refs.~\cite{Motl1}, \cite{Motl2} and \cite{Andersson}.  We have found no way of solving this problem without the presence of such anti-Stokes lines.     
\ignore{One possibility would be to compare the monodromies at the event horizon and infinity.  We can do this by starting at point $g'$ and following the anti-Stokes lines through $f$ and $g$ while we apply `Stokes phenomenon''\cite{Stokes} as we cross the Stokes lines and then loop around the horizon in clockwise direction returning back to $g'$.  Now the final result needs to match the monodromy at the horizon which gives us one equation.  Then we can do the same loop this time starting at $g'$ and going through $e$ instead of $f$ and then loop around the horizon in clockwise direction.  This result should be matched to the monodromy at infinity that gives us a second equation.  We have two equations which can be solved but it turns out that the real part of $\omega$ is zero. This cannot be right since from spherically symmetric black holes we know that Re$(\omega)\propto \ln (3)$. Thus, application of the method in this parametrization requires considerably more care.}

In order to overcome this problem we use a change of variable of the form
\beeq
{x}=y^q~.
\label{new-y}
\eneq 
In this new coordinate, the tortoise-like (spatial) coordinate is defined by
\beeq
dz=\frac{dy}{\bar{f}(y)} ~,
\label{zy}
\eneq
where 
\beeq
\bar{f}(y)=\frac{y^{q(1-b)}-2GM}{qy^{q-1}} ~.
\label{fy}
\eneq
The new horizon is located at  
\beeq
y_h=(2GM)^{1/q(1-b)} ~.
\label{newhorizon}
\eneq  
\ignore{The modified version of the wave equation (\ref{schrodinger1}) becomes 
\beeq
\frac{d^2\Psi}{dy^2}+R(y)\Psi=0 ~,
\label{schrodinger-y}
\eneq
where $\Psi = \sqrt{f(y)/(qy^{q-1})}~\bar{\Psi}$ and 
\begin{eqnarray}
R(y)&=&\left(f\over {qy^{q-1}}\right)^{-2} \nonumber \\
&\cdot&\left\{ \omega^2-U(y)+\frac{1}{4}\left[\frac{d}{dy}\left(f\over {qy^{q-1}}\right)\right]^2 
-\frac{1}{2}\left(f\over {qy^{q-1}}\right)\frac{d^2}{dy^2}\left(f\over {qy^{q-1}}\right)\right\}~.
\label{Ry}
\end{eqnarray}
}
Since $\bar{f}(y) \rightarrow y^{1-q}/q$ as $y \rightarrow 0$, and $h=y^{qa}$ in this new coordinate, the behavior of the function $Q^2$ close to the origin of the complex $y$-plane can be derived by simply substituting $1-\beta=q$, and $\alpha=qa$ in Eq. (\ref{Q^2-general}), where we get
\beeq
Q^2 = R-{1 \over 4y^2} \approx \frac{\omega^2}{\bar{f}^2}-\frac{q^2(a-1)^2}{4y^{2}}~.
\label{newQ^2}
\eneq
We now have the freedom to increase the number of zeros of the function $Q^2$ with the help of our arbitrary parameter $q$.  The important point is to choose $q$ in such a way that we have enough number of zeros to accommodate not only for the anti-Stokes line encircling the event horizon on the positive real axis, but also for anti-Stokes lines extending to infinity in the complex $y$-plane.  
For a particular value of $b$ in the region $0<b<1$, there may be more than one choice of the parameter $q$ that produces the desired behavior of anti-Stokes lines.  To make matters concise however, let us assume that $b$ is a rational number which may be written as the ratio of two integers
\begin{figure}[tb]
\begin{center}
\includegraphics[height=5.5cm]{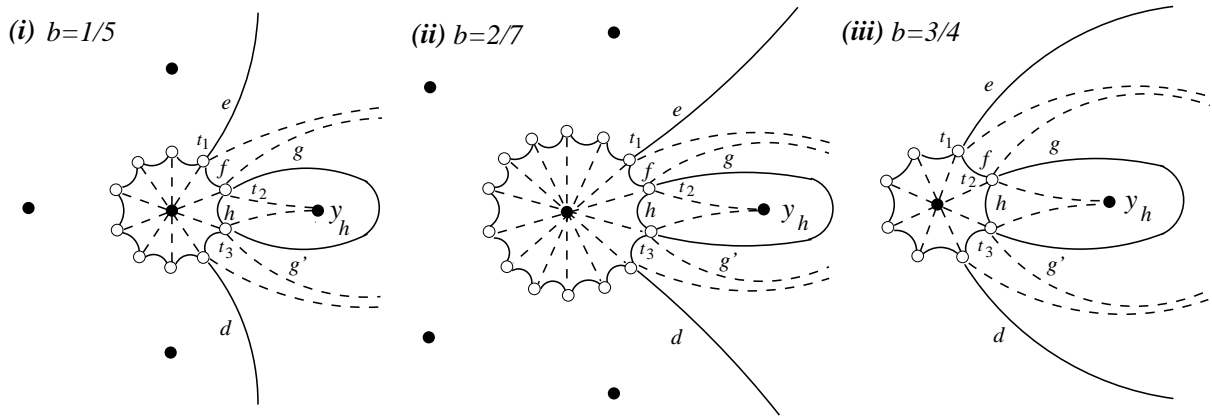}
\end{center}
\caption{A schematic illustration of Stokes (dashed) and anti-Stokes (solid) lines for $b=1/5$, $2/7$, and $3/4$ with $q=5$, $7$, and $4$ respectively in the complex $y$-plane.  The hollow circles represent the zeros of $Q^2$, while the filled circles are the poles at the origin, the event horizon, and the fictitious horizons.  For $\omega = \omega_R+i\omega_I$, with $\omega_R > 0$ and  $\omega_I < 0$, the anti-Stokes line labeled $e$ asymptotes to a straight line at an angle $\arctan(-\omega_I/\omega_R)/qb$.}
\label{schem-b}
\end{figure}
\beeq
b=\frac{m}{n}~,
\label{b}
\eneq
where $m<n$.  We then conjecture that the choice of $q=n$ will always produce the desired structure of anti-Stokes lines for any rational $b$ in the interval $0<b<1$.  The behavior of Stokes and anti-Stokes lines are schematically plotted in Fig.~\ref{schem-b} for particular values of $b=1/5$, $2/7$, and $3/4$ with $q=5$, $7$, and $4$ respectively.  
Note that in addition to the event horizon on the positive real axis, we may also have extra fictitious horizons in the complex $y$-plane.  The horizons are located at
\beeq
y_k=e^{i2k\pi/q(1-b)}|y_h|=e^{i2k\pi/(n-m)}|y_h|~,
\label{fictitious}
\eneq
where $k=0,1,2,...,n-m-1$ with $k=0$ corresponding to the event horizon.  Since the number of zeros of the function $Q^2$ is $2q=2n$, the anti-Stokes lines labeled $e$ and $d$ in Fig.~\ref{schem-b} emanate with an angle $3\pi/n$ with respect to each other. In addition, according to Eq. (\ref{fictitious}), the argument of the first fictitious horizon in the complex $y$-plane is $2\pi/(n-m)$.  This means that the angle between the two fictitious horizons on either side of the event horizon is $4\pi/(n-m)$.  Based on our considerations $3\pi/n < 4\pi/(n-m)$.  This means that our fictitious horizons will always stay to the left of the anti-Stokes lines labeled $d$ and $e$ in Fig. \ref{schem-b}.

The behavior of the Stokes lines on which Re$(\int Q dy)=0$ and anti-Stokes lines on which Im$(\int Qdy)=0$ near the origin of the complex $y$-plane may be determined using Eq. (\ref{newQ^2}).  Let us investigate the behavior of these lines away from the origin.  In this region we may write
\beeq
Q^2 \sim  \left( \frac{qy^{q-1}\omega}{y^{q(1-b)}-2GM}\right)^2=\left( \frac{ny^{n-1}\omega}{y^{n-m}-2GM}\right)^2~,
\label{Q^2large-y}
\eneq 
when we take $b=m/n$ and $q=n$.  To determine the behavior of the Stokes and anti-Stokes lines we need to evaluate $\int Qdy$.  Integrating $Q$ with respect to $y$ gives
\beeq
\int Q dy \sim \pm \left\{ \frac{n}{m}y^m+\frac{n}{n-m}\sum_{k=0}^{n-m-1} y_k \left[ \sum_{p=0}^{m-2} (y_k)^p\frac{y^{m-1-p}}{m-1-p}+(y_k)^{m-1}\ln\left(1-\frac{y}{y_k}\right)\right]\right\}\omega ~,
\label{Qdy}
\eneq 
where we added a constant to make sure that $\int Q dy=0$ at $y=0$.  It is clear that $\int Qdy$ is a multivalued function.  To make this function single valued we need to introduce branch cuts in the complex $y$-plane emanating from each of the horizons.  Near the horizons this equation may be simplified to give
\beeq
\int Q dy \sim \pm  \left[\frac{n}{n-m} y_k^{m} \ln\left(1-\frac{y}{y_k}\right)\right]\omega~.
\label{Qdy2}
\eneq 
Assuming $\omega$ to be purely imaginary, it is easy to see that Im($\int Qdy$) is single valued around $y_k$ with $k=0$ and $k=(n-m)/2$ (if $n-m$ is even), but around all the other fictitious horizons, where $y_k$ is complex, Im($\int Qdy$) is multivalued.  This means that anti-Stokes lines, on which Im$(\int Qdy)=0$, are well defined around the horizons located on the real axis and therefore are allowed to cross the branch cuts emanating from these horizons.  Around all the other fictitious horizons the anti-Stokes lines are not well defined and therefore are not allowed to cross the branch cuts.  Some of the anti-Stokes lines may terminate on these fictitious horizons or on the branch cuts emanating from them.  It is also easy to see that Re$(\int Qdy)$ is multivalued around all the horizons except around $y_k$ with $k=(n-m)/4$ and $k=3(n-m)/4$ (if $n-m$ is even).  Therefore Stokes lines, on which Re$(\int Qdy)=0$, are not allowed to cross the branch cuts emanating from the horizons which are not located on the imaginary axis.  

\begin{figure}[tb]
\begin{center}
\includegraphics[height=9.0cm]{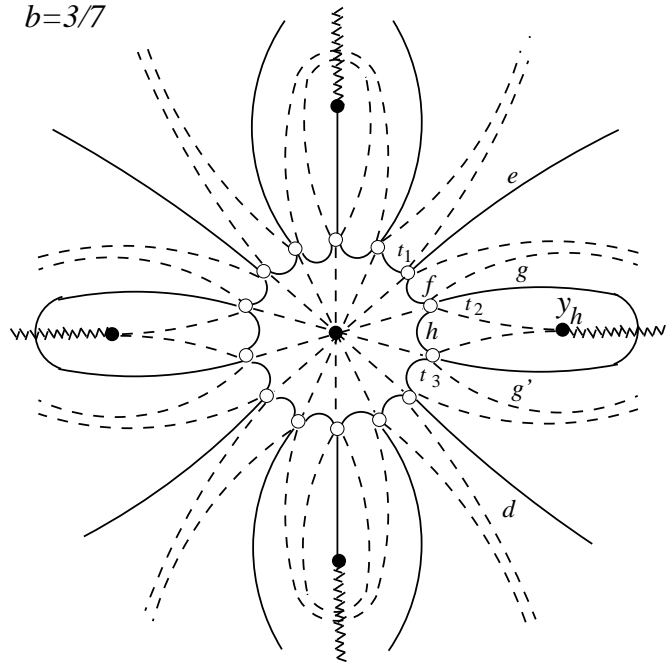}
\end{center}
\caption{A schematic illustration of Stokes (dashed) and anti-Stokes (solid) lines for $b=3/7$ with $q=7$ in the complex $y$-plane.  The hollow circles represent the zeros of $Q^2$, while the filled circles are the poles at the origin and the horizons.  The branch cuts emanating from the horizons are illustrated with zigzagged lines.}
\label{schem-b3-7}
\end{figure}

As we mentioned earlier, the fictitious horizons are always located to the left of the unbounded anti-Stokes lines $d$ and $e$.  Therefore we are able to introduce the branch cuts emanating from the fictitious horizons in a way that they do not affect the behavior of the Stokes and anti-Stokes lines on the right hand side of the complex $y$-plane.  We will see in the next section that the behavior of the Stokes and anti-Stokes lines to the left of the lines $d$ and $e$ are irrelevant in our calculations and therefore have not been plotted in Fig. \ref{schem-b}.  For the interested reader we plot the behavior of the Stokes and anti-Stokes lines on the entire complex $y$-plane for $b=3/7$ with $q=7$ in Fig. \ref{schem-b3-7}.  Note that in this figure two of the anti-Stokes lines terminate on the fictitious horizons located on the imaginary axis.  For more examples on the behavior of anti-Stokes lines in the vicinity of the fictitious horizons refer to Ref.~\cite{Natario}.  

\sxn{Quasinormal mode frequencies}

In the previous section we conjectured that with the choice of $q=n$ we may always have a desired structure of Stokes and anti-Stokes lines similar to the diagrams shown in Fig.~\ref{schem-b} for any rational $b=m/n$ .  In this section we will show how to solve for the highly damped QNM frequencies using the diagrams in Fig.~\ref{schem-b}.  

We start at point $e$  on the anti-Stokes line extending to infinity with an outgoing wave solution
\beeq
\Psi_e=\Psi_1^{(t_1)}~.
\label{Psie}
\eneq
Note that in order to have the outgoing-wave solution at infinity to be proportional to $\Psi_1$ and the ingoing-wave solution at the event horizon to be proportional to $\Psi_2$, we have chosen the phase of the square-root of $Q^2$ such that $Q=\sqrt{R-1/4y^2}\sim qy^{qb-1}\omega$ as $y\rightarrow \infty$.  To do this, we need to introduce branch cuts emanating from the simple zeros of $Q^2$.  These branch cuts may usually be placed in such a way that they do not affect our analysis.  

To go from $e$ to $f$ we cross a Stokes line emanating from $t_1$ on which $\Psi_1$ is dominant according to our choice of phase for $Q$.  When we cross a Stokes line we need to account for ``Stokes phenomenon''\cite{Stokes} which means that the coefficient of the dominant solution ($\Psi_1$ in this case) on the relevant Stokes line  remains unchanged and the coefficient of the subdominant solution ($\Psi_2$ in this case) picks up a contribution proportional to the coefficient of the dominant solution.  The constant of proportionality is called ``Stokes constant''.  In our case where we are dealing with isolated simple zeros, the Stokes constant is simply $+i$ for crossing in the positive (anti-clockwise) direction, and $-i$ for crossing in the negative (clockwise) direction.  Therefore, after crossing the first Stokes line, our solution becomes
\beeq
\Psi_f=\Psi_1^{(t_1)}-i\Psi_2^{(t_1)}~.
\label{Psif}
\eneq
This solution will not change in character when we move along the anti-Stokes line labeled $f$ from $t_1$ to $t_2$ because no Stokes lines cross this contour, but we need to change the lower limit of the phase-integral of $\Psi_1$ and $\Psi_2$ to $t_2$.  This means that we need to evaluate the integrals of the type
\beeq
\gamma_{ij} =\int_{t_i}^{t_j} Q dy \approx  \int_{t_i}^{t_j} \left[\frac{q^2y^{2(q-1)}}{(2GM)^2}\omega^2-\frac{q^2(a-1)^2}{4y^{2}}\right]^{1/2}dy ~,
\label{gamma1}
\eneq
where $t_i$ and $t_j$ are two simple zeros of the function $Q^2$.
\ignore{Note that this integral is nonzero only when $2q=${\it even number}.  Therefore $q$ needs to be an integer greater than one; i.e.
\beeq
q=2, 3, 4, ...~.
\label{q}
\eneq  
}
We can solve this integral by introducing a new variable $\eta=2\omega y^q/[2GM(a-1)]$ which maps the zeros to $-1$ or $+1$, and this integral becomes
\beeq
\gamma_{ij}= \frac{\sqrt{a(a-2)+1}}{2} \int_{\mp1}^{\pm 1} \left(1-\frac{1}{\eta^2}\right)^{1/2}d\eta=\mp \frac{a-1}{2}\pi~.
\label{gamma2}
\eneq

We now can change the lower limit of the phase integral in (\ref{Psif}) to $t_2$ by simply writing
\beeq
\Psi_f=e^{i\gamma_{12}}\Psi_{1}^{(t_2)}-ie^{-i\gamma_{12}}\Psi_{2}^{(t_2)}~.
\label{Psif1}
\eneq
We may extend this solution to $g$ by crossing another Stokes line on which $\Psi_1$ is dominant.  Hence, at $g$ we obtain
\beeq
\Psi_g=e^{i\gamma_{12}}\Psi_{1}^{(t_2)}-i\left(e^{i\gamma_{12}}+e^{-i\gamma_{12}}\right)\Psi_{2}^{(t_2)}~.
\label{Psig}
\eneq

We then proceed from $g$ to $g'$ on the anti-Stokes line that loops around the pole at $y_h$.  This contour does not cross any Stokes line, but we need to change the lower limit of the phase integral to $t_3$.  To do this, we may write 
\beeq
\Psi_{g'}=e^{i\gamma_{12}}e^{i\tilde{\gamma}_{23}}\Psi_{1}^{(t_3)}-i\left(e^{i\gamma_{12}}+e^{-i\gamma_{12}}\right)e^{-i\tilde{\gamma}_{23}}\Psi_{2}^{(t_3)}~,
\label{Psig'}
\eneq
where $\tilde{\gamma}_{23}=\Gamma+\gamma_{23}$.  Here $\Gamma$ is the integral of $Q$ along a contour encircling the pole at $y_h$ in the clockwise direction and may be evaluated using the fact that
\beeq
\Gamma=\oint Q dy= -2\pi i \mathop{Res}_{y=|y_h|} Q= -2\pi i\left[{\frac{1}{1-b}(2GM)^{b/(1-b)}\omega}\right] ~.
\label{Gamma}
\eneq
In terms of the generic black hole Hawking temperature
\beeq
T_{BH}={\kappa\over 2\pi}=\left.{1\over 4\pi}{df\over dx}\right|_{x=x_h}~,
\label{temperature}
\eneq
$\Gamma$ is therefore
\beeq
\Gamma = - i\frac{\omega}{2T_{BH}}~.
\label{Gamma1}
\eneq

We now connect the solution to the point $h$ by stepping inside the anti-Stokes line encircling the horizon.  We cross the first Stokes line on which $\Psi_2$ is dominant.  To account for the Stokes phenomenon we simply replace $\Psi_2$ with $\Psi_2+i\Psi_1$.  Then we change the lower limit of the phase-integral to $t_2$ and we replace $\Psi_2$ with $\Psi_2+i\Psi_1$ again to cross the second Stokes line on which $\Psi_2$ is dominant.  Finally we return back to $g$ where
\ignore{
\beeq
\Psi_h=\left[e^{i\gamma_{12}}e^{i\Gamma}+\left(e^{i\gamma_{12}}+e^{-i\gamma_{12}}\right)e^{2i\gamma_{32}}e^{-i\Gamma}\right]\Psi_{1}^{(t_2)}-i\left(e^{i\gamma_{12}}+e^{-i\gamma_{12}}\right)e^{-i\Gamma}\Psi_{2}^{(t_2)}~.
\label{Psid}
\eneq
After crossing the second Stokes line on which $\Psi_2$ is dominant, we return back to $g$ again where the solution is
}
\begin{eqnarray}
\Psi_{\bar g}&=&\left[e^{i\gamma_{12}}e^{i\Gamma}+\left(e^{i\gamma_{12}}+e^{-i\gamma_{12}}\right)\left(1+e^{2i\gamma_{32}}\right)e^{-i\Gamma}\right]\Psi_{1}^{(t_2)}\nonumber\\
&-&i\left(e^{i\gamma_{12}}+e^{-i\gamma_{12}}\right)e^{-i\Gamma}\Psi_{2}^{(t_2)}~.
\label{Psigbar}
\end{eqnarray}
The bar on $g$ is to distinguish this solution from $\Psi_g$.

In order to return back to $e$, we reverse our initial steps.  We first go back to $f$ by crossing the first Stokes line.  Then we change the lower limit of the phase-integral to $t_1$ and we cross the next Stokes line.  Our final solution using the fact that $\gamma_{12}=\gamma_{32}$ is   
\ignore{At $t_1$ the solution becomes
\beeq
\Psi_{\bar b}=\left[e^{i\Gamma}+\left(1+e^{-2i\gamma_{12}}\right)\left(1+e^{2i\gamma_{32}}\right)e^{-i\Gamma}\right]\Psi_{1}^{(t_2)}+i\left[e^{2i\gamma_{12}}e^{i\Gamma}+\left(e^{2i\gamma_{12}}+1\right)e^{2i\gamma_{32}}e^{-i\Gamma}\right]\Psi_{2}^{(t_2)}~.
\label{Psibbar}
\eneq
}
\begin{eqnarray}
\Psi_{\bar e}&=&\left[e^{i\Gamma}+\left(2+e^{2i\gamma_{12}}+e^{-2i\gamma_{12}}\right)e^{-i\Gamma}\right]\Psi_{1}^{(t_2)} \nonumber\\
&+&i\left[\left(1+e^{2i\gamma_{12}}\right)e^{i\Gamma}+\left( 2+2e^{2i\gamma_{12}}+e^{-2i\gamma_{12}}+e^{4i\gamma_{12}}\right)e^{-i\Gamma}\right]\Psi_{2}^{(t_2)}~.
\label{Psiebar}
\end{eqnarray}

Comparing $\Psi_e$ and $\Psi_{\bar{e}}$ requires that
\beeq
\left(1+e^{2i\gamma_{12}}\right)e^{i\Gamma}+\left( 2+2e^{2i\gamma_{12}}+e^{-2i\gamma_{12}}+e^{4i\gamma_{12}}\right)e^{-i\Gamma}=0~.
\label{WKBcondition}
\eneq
This will lead to 
\beeq
\Psi_{\bar{e}}=e^{-i\Gamma}\Psi_e~,
\label{monodramy}
\eneq
which gives us the correct monodramy in the clockwise direction of the form $e^{-i\Gamma}$.  Equation (\ref{WKBcondition}) gives us the WKB condition 
\beeq
e^{2i\Gamma}=-1-2\cos {2\gamma_{12}}~.
\label{WKBcondition1}
\eneq
Substituting the values of $\gamma_{12}$ and $\Gamma$ from Eqs. (\ref{gamma2}) and (\ref{Gamma1}) respectively, the general result is
\beeq
\frac{\omega}{T_{BH}} = (2n+1)\pi i+\ln[-1-2\cos((a-1)\pi)]~.
\label{omega_final}
\eneq
Note that this result is independent of the choice of the parameters $b$ and $q$. However, the result is only rigorous for rational $b$. Nonetheless, this result a common result for the generic class of black holes studied in this paper.
As we mentioned earlier, we are particularly interested in the physically motivated case where $a=1$.  For all odd integer $a$, including $a=1$, we get the familiar WKB condition
\beeq
\frac{\omega}{T_{BH}} = (2n+1)\pi i + ln(3)~.
\label{WKBcondition2}
\eneq
For even $a$, on the other hand, the imaginary part is zero and there appear to be no large damping QNMs.
It is important to realize in the context of 2-d dilaton gravity, the parameter $a$ can in principle take on any non-negative real value, in which case the real part of the large damping QNM frequency is not the logarithm of an integer.
\begin{figure}[tb]
\begin{center}
\includegraphics[height=7.0cm]{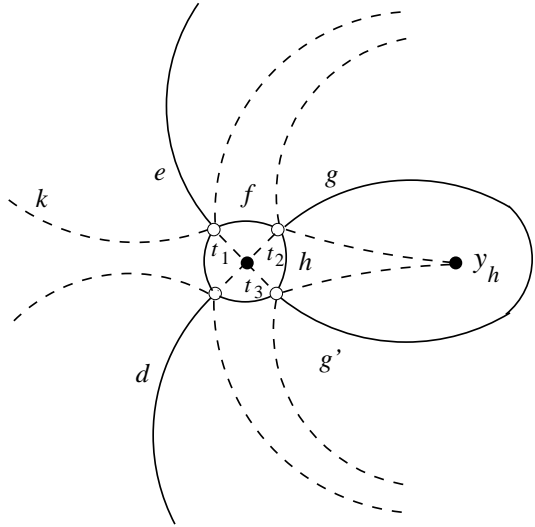}
\end{center}
\caption{A schematic illustration of Stokes (dashed) and anti-Stokes (solid) lines for $1/2 \leq b <1$ with $q=2$ in the complex $y$-plane.  The hollow circles represent the zeros of $Q^2$, while the filled circles are the poles at the origin and  the event horizon.  For $\omega = \omega_R+i\omega_I$, with $\omega_R > 0$ and  $\omega_I < 0$, the anti-Stokes line labeled $e$ asymptotes to a straight line at an angle $\arctan(-\omega_I/\omega_R)/qb$ and the Stokes line labeled $k$ asymptotes to a straight line at an angle $[\pi/2+\arctan(-\omega_I/\omega_R)]/qb$.}
\label{schem-q2}
\end{figure} 

\ignore{As we mentioned earlier, in this paper we are particularly interested in the physically motivated case where $a=1$.  Note that when $a=1$, the zeros of $Q^2$ shrink to the origin of the complex plane.  Therefore we need to keep $a$ an arbitrary (coupling) parameter during the calculations and let 
$a \rightarrow 1$ after we get the final answer.  The same thing arises in the calculations of the QNM frequencies of scalar perturbations in Schwarzschild black holes.  Note that taking $j=0$ for scalar perturbations early in the calculations will result in a wrong result.}

\sxn{A different choice of the parameter $q$}

So far we have shown how to solve for the highly damped QNM frequencies using the change of variable (\ref{new-y}) with the particular choice of $q=n$, where $n$ is the denominator of the rational parameter $b$.  We mentioned earlier that there may be more than one choice of the parameter $q$ for any particular value of $b$, which may produce the desired behavior of anti-Stokes lines of the type shown in Fig.~\ref{schem-b}.  In this section we would like to explore a different choice of the parameter $q$.  
If we restrict ourselves to the region $1/2 \leq b <1$, it turns out that with the choice of $q=2$ we always have $q(1-b)\leq 1$.  This means that according to Eq. (\ref{newhorizon}) we always have only one horizon located on the positive real axis with no fictitious horizons in the complex $y$-plane.  For this choice of $q$, according to Eq. (\ref{newQ^2}), the function $Q^2$ has four simple zeros. The behavior of the Stokes and anti-Stokes lines in this region of $b$ is shown in Fig.~\ref{schem-q2}.  Note that the choice of $q=2$ is independent of the parameter $b$.  In other words, in this range of $b$ we do not need to assume that $b$ is a rational number.  Therefore, our analysis in the region $1/2\leq b<1$ include both rational and irrational $b$.  Using the same methodology as in the previous section we can solve for the highly damped QNM frequencies, and the result will be the same as in Eq. (\ref{omega_final}).

\ignore{Let us take $q=1/(1-b)$.  The advantage of this choice is that we always have only one horizon on the positive real axis.  If $b$ is a rational number then we can write
\beeq
b=\frac{m}{n}~,
\label{b}
\eneq
where $m$ and $n$ are integer numbers and since $0<b<1$ then $m<n$.  If $n-m$ is an odd number then the number of zeros is equal to $2n$.  Since $n$ cannot be less than $2$, therefore $2n \geq 4$.  This means that we always have more than two anti-Stokes lines which will go to infinity in the complex x-plane and the problem can be solved.  This mathematical proof includes a wide range of black holes including Schwarzschild black holes with even number of dimensions because our two dimensional dilaton metric can be reduced to Schwarzschild metric by taking $m=1$ and $n=d-2$ where $d$ is the dimension of the black hole.  The situation changes if $n-m$ is an even number.  In other words both $n$ and $m$ are odd numbers.  In this case the number of zeros equal to $n$ which is an odd number.  The only way we found to overcome this problem is taking $q=n$.  In this case the number of horizons is equal to $n-m$ which is an even number and the number of zeros will be equal to $2n$.  Now we have even number of zeros and the ratio between the number of zeros and the number of horizons is $2n/(n-m)$.  For any particular $n$ this ratio has a minimum when $m=1$.  Taking $m=1$, this ratio approaches to 2 for large $n$. Therefore the number of zeros is always more than two times the number of horizons.  For these cases we have no proof that the solution exist for all cases in this family, but we conjecture that for every case when $n-m=$ is an odd number the structure of the anti-Stokes lines on the right side of the complex y-plane will always stay the same.  That is we always have an anti-Stokes line which loops around the horizon on the positive real axis and two anti-Stokes lines which extend to infinity from two zeros in the vicinity of the loop.  A subgroup of these type of black holes in which $n-m$ is a odd number is the Schwarzschild black holes with odd number of dimensions.
}

\sxn{Summary and discussion}

In Sections 4 and 5 we rigorously showed that for a class of 2-d dilaton gravities the highly damped QNM spectrum for single horizon black holes is given by:
\beeq
\frac{\omega}{T_{BH}} = (2n+1)\pi i+\ln[-1-2\cos((a-1)\pi)]~,
\label{omega_final2}
\eneq
where $a$ is the exponent of the  coupling $h(\phi)$ near the origin, in tortoise coordinates. This class of theories includes, among others, the higher dimensional Schwarzschild solutions.
 
Although the detailed calculations in Sections 4 and 5 dealt with a specific class of models, it is now straightforward to understand the general conditions under which the final result (\ref{omega_final2}) applies. The general form of $Q^2$ is given in Eq.~(\ref{Q^2-general}). The number of anti-Stokes lines emerging from the origin and their angle are determined by the behavior of $Q^2$ near the origin. $F\to x^\beta$, and one is free to choose a coordinate system in which $1-\beta$ is an integer greater than $1$. It is clear from (\ref{Q^2-general}) that in such a coordinate system there will be $2(1-\beta)$ zeros of $Q^2$ and hence $2(1-\beta)$ anti-Stokes lines emerging from the origin, separated by angles of $2\pi/2(1-\beta)$. Assuming that $F(x)$ has a simple zero at the horizon on the real axis, then the two anti-Stokes lines closest to the real axis can be expected to meet on the real axis to the right of the horizon, as in the above figures. As long as there are no solutions to $F(x)=0$ on the complex plane close to the real axis, the next closest anti-Stokes lines will in fact escape to infinity, and the contour that we have used will yield precisely the condition in Eq.~(\ref{omega_final}). 

Note that this analysis immediately justifies the conclusions in Refs.~\cite{Gabor3} and \cite{Das1}, which used the monodromy arguments of Motl\cite{Motl1, Motl2} to calculate the change of phase in tortoise coordinates as the solution circled the origin. These conclusions assume that in tortoise coordinates one must use a contour that encircles the origin by an angle of $3\pi$. This angle, however, must be justified in terms of anti-Stokes lines in coordinate $x$. We now see that the relevant anti-Stokes lines in the $x$-plane are the second ones on either side of the positive real axis, and that they are  generically separated by an angle of
\beeq
\Delta\theta = 3 {2\pi\over 2(1-\beta)}~.
\eneq
When $F\sim x^\beta$ the relationship between the tortoise coordinate and our $x$-coordinate is $z\sim x^{1-\beta}$, so the corresponding angle in tortoise coordinates is $(1-\beta)\Delta \theta = 3 \pi$. Thus the results are completely general and valid.

The general consensus among researchers seems to be that the large damping black hole QNM frequency has little to do with the quantum gravitational microstates that account for black hole entropy.  The results in the present paper seem to support this view: the coefficient of the highly damped QNM frequency derives from the exponent that determines the $x\to zero$ limit of the matter-gravity coupling, expressed in tortoise coordinates. In the general class of 2-d dilaton gravity models this exponent can take on any non-negative value, so that the coefficient that relates frequency to temperature is only $\ln(3)$ in very special cases (i.e. odd integers). For spherically symmetric, higher dimensional black holes, the $\ln(3)$ seems to be more universal, but again, it is difficult to see how it may be connected to the quantum gravitational microstates of the horizon.

Moreover, if the connection between highly damped QNMs and quantum gravitational microstates were real, it should persist in some form for all black holes, not just single horizon, asymptotically flat ones. This seems not to be the case. In fact, it seems that adding even an infinitesimal amount of charge, for example, so that a second horizon appears near the origin, changes qualitatively the nature of the highly damped quasinormal modes. An investigation of this mechanism in the general framework presented here is currently in progress.

\vskip .5cm

\leftline{\bf Acknowledgments}
 We are grateful to Joey Medved, Saurya Das and Brian Dolan for useful discussions. We also thank J. Natario and R. Schiappa for a private communication in which they stressed the importance of the asymptotic properties of the anti-Stokes lines. This research was supported in part by the Natural Sciences and Engineering Research Council of Canada.


\def\jnl#1#2#3#4{{#1}{\bf #2} (#4) #3}

\def\Zphys{{\em Z.\ Phys.} }
\def\jssc{{\em J.\ Solid State Chem.\ }}
\def\jpsJ{{\em J.\ Phys.\ Soc.\ Japan }}
\def\ptps{{\em Prog.\ Theoret.\ Phys.\ Suppl.\ }}
\def\PTP{{\em Prog.\ Theoret.\ Phys.\  }}
\def\LNC{{\em Lett.\ Nuovo.\ Cim.\  }}

\def\JMP{{\em J. Math.\ Phys.} }
\def\NPB{{\em Nucl.\ Phys.} B}
\def\NP{{\em Nucl.\ Phys.} }
\def\PLB{{\em Phys.\ Lett.} B}
\def\PL{{\em Phys.\ Lett.} }
\def\PRL{\em Phys.\ Rev.\ Lett. }
\def\PRB{{\em Phys.\ Rev.} B}
\def\PRD{{\em Phys.\ Rev.} D}
\def\PR{{\em Phys.\ Rev.} }
\def\PRe{{\em Phys.\ Rep.} }
\def\AP{{\em Ann.\ Phys.\ (N.Y.)} }
\def\RMP{{\em Rev.\ Mod.\ Phys.} }
\def\ZPC{{\em Z.\ Phys.} C}
\def\SCI{\em Science}
\def\CMP{\em Comm.\ Math.\ Phys. }
\def\MPLA{{\em Mod.\ Phys.\ Lett.} A}
\def\IJMPB{{\em Int.\ J.\ Mod.\ Phys.} B}
\def\cmp{{\em Com.\ Math.\ Phys.}}
\def\JPA{{\em J.\  Phys.} A}
\def\CQG{\em Class.\ Quant.\ Grav.~}
\def\ATMP{\em Adv.\ Theoret.\ Math.\ Phys.~}
\def\PRSA{{\em Proc.\ Roy.\ Soc.} A }
\def\ibid{{\em ibid.} }
\vskip 1cm

\leftline{\bf References}

\renewenvironment{thebibliography}[1]
        {\begin{list}{[$\,$\arabic{enumi}$\,$]}  
        {\usecounter{enumi}\setlength{\parsep}{0pt}
         \setlength{\itemsep}{0pt}  \renewcommand{\baselinestretch}{1.2}
         \settowidth
        {\labelwidth}{#1 ~ ~}\sloppy}}{\end{list}}


\end{document}